\documentclass[orivec]{llncs}
\usepackage[toc,title,page]{appendix}
\usepackage[dvips]{graphicx}
\usepackage{amsmath,amssymb,bm,fancybox}
\usepackage{bm}
\usepackage{cite}
\usepackage{verbatim} 
\usepackage{url}
\usepackage[dvipdfmx]{color}
\usepackage{setspace}
\usepackage{cite}
\usepackage{here}
\usepackage{wrapfig}

\usepackage{tikz} 
\usetikzlibrary{calc}
\usetikzlibrary{positioning}
\usetikzlibrary{shapes,arrows}

\newcommand{\bin}{\{0,1\}}

\newcommand{\ra}{\rightarrow}

\newcommand{\ol}[1]{\overline{#1}}

\newlength{\ralen}
\def\C{\clubsuit}
\def\H{\heartsuit}

\newcommand{\back}{\raisebox{\ralen}{\framebox(11,13){\bf ?}}}

\tikzset{vtx/.style={shape=circle, draw, outer sep=2pt}}

\date{}

\begin{document}

\title{A Note on Single-Cut Full-Open Protocols}

\author{
  Kazumasa Shinagawa\inst{1,3}
  \and
  Koji Nuida\inst{2,3}
}
\institute{
  {University of Tsukuba, Ibaraki, Japan}\\
  \email{shinagawa@cs.tsukuba.ac.jp}
  \and
  {Institute of Mathematics for Industry, Kyushu University, Fukuoka, Japan}
  \and
  {National Institute of Advanced Industrial Science and Technology, Tokyo, Japan}
}

\maketitle

\begin{abstract}
Card-based cryptography is a research area that realizes cryptographic protocols such as secure computation by applying shuffles to sequences of cards that encode input values. A single-cut full-open protocol is one that obtains an output value by applying a random cut to an input sequence of cards, after which all cards are opened. In this paper, we propose three single-cut full-open protocols: two protocols for three-variable functions and one protocol for a four-variable function.
\end{abstract}

\section{Introduction}\label{sec:introduction}

Cryptographic primitives such as secure computation and zero-knowledge proofs can be realized using physical playing cards, and the study of constructing such protocols is referred to as \emph{card-based cryptography}~\cite{BoerEC1989,KilianC1994,MizukiIJIS2014}. 

The first card-based protocol was the so-called \emph{five-card trick} proposed by den Boer~\cite{BoerEC1989}, which securely computes the logical AND value $xy$ of two input bits $x,y \in \bin$.
A notable feature of this protocol is that it applies a single shuffle, known as a \emph{random cut}, and subsequently opens all the cards at the end of the execution. Protocols with this property are referred to as \emph{single-cut full-open (SCFO) protocols}~\cite{ShinagawaICISC2018,ShinagawaFCT2025}\footnote{This notion was first introduced by Shinagawa--Mizuki~\cite{ShinagawaICISC2018}, who referred to such protocols as \emph{single-cut garbage-free protocols}, focusing on the property that no face-down cards (i.e., garbage) remain at the end of the protocol. The name of \lq\lq single-cut full-open protocols\rq\rq{} was first used by Shinagawa--Nuida~\cite{ShinagawaFCT2025}. A generalized notion \lq\lq single-shuffle full-open protocols\rq\rq{}, which are full-open protocols with any type of a single shuffle, was introduced by Shinagawa--Nuida~\cite{ShinagawaEPRINT2022,ShinagawaSTACS2025}.}.

A random cut is an operation that applies a cyclic shift by a uniformly random number to a sequence of cards. When a random cut is applied to a sequence of $n$ face-down cards, denoted by $\langle\,\back\,\back\,\cdots\,\back\,\rangle$, each of the $n$ possible sequence is equally likely to occur. Moreover, which sequence is actually chosen must remain hidden from all players. 

In general, if any face-down cards remain at the end of a protocol, extra effort may be required to clean them up or reuse them, often necessitating additional shuffles. By contrast, when the protocol satisfies the full-open property, such additional operations become unnecessary. Moreover, random cuts are known to be the easiest shuffle to implement. Therefore, it is fair to say that SCFO protocols represent the simplest class of card-based protocols. 

This paper is organized as follows: In Section \ref{s:known}, we review existing SCFO protocols, and in Section \ref{s:our}, we propose new SCFO protocols for three functions.

\section{Known Protocols}\label{s:known}

\subsection{Protocol for $x\oplus y$}

This protocol was proposed by Shinagawa--Mizuki~\cite{ShinagawaICISC2018}.

\begin{enumerate}
    \item The input sequence is given as follows:
    \[
    \underset{x}{\back} \, \underset{\ol{x}}{\back} \, \underset{y}{\back}\, \underset{\ol{y}}{\back} \,.
    \]
    \item Apply a random cut to the sequence as follows: 
    \[
    \left\langle \; \back\,\back\,\back\,\back \;\right\rangle. 
    \]
    \item Open all cards. Output $0$ if it is a cyclic shift of $\H\C\H\C$, and $1$ if it is a cyclic shift of $\H\H\C\C$.
\end{enumerate}

\subsection{Protocol for $xy$ (Five-Card Trick)}

This protocol was proposed by den Boer~\cite{BoerEC1989}.

\begin{enumerate}
    \item The input sequence is given as follows:
    \[
    \underset{\ol{x}}{\back} \, \underset{x}{\back} \, \underset{\heartsuit}{\back} \, \underset{y}{\back}\, \underset{\ol{y}}{\back} \,.
    \]
    \item Apply a random cut to the sequence as follows: 
    \[
    \left\langle \; \back\,\back\,\back\,\back\,\back \;\right\rangle. 
    \]
    \item Open all cards. Output $0$ if it is a cyclic shift of $\H\C\H\C\H$, and $1$ if it is a cyclic shift of $\H\H\H\C\C$.
\end{enumerate}

\subsection{Protocol for $(x=y=z)?$ (Six-Card Trick)}

This protocol was proposed by den Heather--Schneider--Teague~\cite{HeatherFAOC2014}. It was independently rediscovered by Shinagawa--Mizuki~\cite{ShinagawaICISC2018} and named it as \emph{six-card trick}. 

\begin{enumerate}
    \item The input sequence is given as follows:
    \[
    \underset{x}{\back} \, \underset{\ol{y}}{\back} \, \underset{z}{\back}\, \underset{\ol{x}}{\back} \, \underset{y}{\back} \, \underset{\ol{z}}{\back} \,.
    \]
    \item Apply a random cut to the sequence as follows: 
    \[
    \left\langle \; \back\,\back\,\back\,\back\,\back\,\back \;\right\rangle. 
    \]
    \item Open all cards. Output $0$ if it is a cyclic shift of $\H\C\H\C\H\C$, and $1$ if it is a cyclic shift of $\H\H\H\C\C\C$. 
\end{enumerate}

\section{Our Protocols}\label{s:our}

In this section, we propose SCFO protocols for three Boolean functions. 

\subsection{Protocol 1: SCFO Protocol for $x \oplus y \oplus z$}

\begin{table}[t]
    \centering
    \begin{tabular}{|ccc|c|c|}\hline
    ~$x$~ & ~$y$~ & ~$z$~ & ~$x \oplus y \oplus z$~ & ~Input Seq.~\\ \hline\hline
    $0$ & $0$ & $0$ & $0$ & ~$\C\C\H\C\C\H\H\H$~ \\
    $0$ & $0$ & $1$ & $1$ & ~$\C\C\H\H\C\H\H\C$~ \\
    $0$ & $1$ & $0$ & $1$ & ~$\C\H\H\C\C\C\H\H$~ \\
    $0$ & $1$ & $1$ & $0$ & ~$\C\H\H\H\C\C\H\C$~ \\
    $1$ & $0$ & $0$ & $1$ & ~$\H\C\C\C\H\H\C\H$~ \\
    $1$ & $0$ & $1$ & $0$ & ~$\H\C\C\H\H\H\C\C$~ \\
    $1$ & $1$ & $0$ & $0$ & ~$\H\H\C\C\H\C\C\H$~ \\
    $1$ & $1$ & $1$ & $1$ & ~$\H\H\C\H\H\C\C\C$~ \\\hline
    \end{tabular}
    \caption{Correctness of Protocol 1}
    \label{tab:XOR}
\end{table}

\begin{enumerate}
    \item The input sequence is given as follows:
    \[
    \underset{x}{\back} \, \underset{y}{\back}\, \underset{\ol{x}}{\back} \, \underset{z}{\back} \, \underset{x}{\back} \, \underset{\ol{y}}{\back}\, \underset{\ol{x}}{\back} \, \underset{\ol{z}}{\back} \,.
    \]
    \item Apply a random cut to the sequence as follows: 
    \[
    \left\langle \; \back\,\back\,\back\,\back\,\back\,\back\,\back\,\back \;\right\rangle. 
    \]
    \item Open all cards. Output $0$ if it is a cyclic shift of $\C\C\H\C\C\H\H\H$, and $1$ if it is a cyclic shift of $\H\H\C\H\H\C\C\C$. 
\end{enumerate}

The correctness of Protocol 1 is shown in Table~\ref{tab:XOR}. 

\subsection{Protocol 2: SCFO Protocol for $\ol{x}y\ol{w} \vee \ol{y}z\ol{w} \vee x\ol{y}w \vee y\ol{z}w$}

\begin{table}[t]
    \centering
    \begin{tabular}{|cccc|c|c|}\hline
    ~$x$~ & ~$y$~ & ~$z$~ & ~$w$~ & ~$\ol{x}y\ol{w} \vee \ol{y}z\ol{w} \vee x\ol{y}w \vee y\ol{z}w$~ & ~Input Seq.~\\ \hline\hline
    $0$ & $0$ & $0$ & $0$ & $0$ & ~$\C\C\C\C\H\H\H\H$~ \\
    $0$ & $0$ & $0$ & $1$ & $0$ & ~$\C\C\C\H\H\H\H\C$~ \\
    $0$ & $0$ & $1$ & $0$ & $1$ & ~$\C\C\H\C\H\H\C\H$~ \\
    $0$ & $0$ & $1$ & $1$ & $0$ & ~$\C\C\H\H\H\H\C\C$~ \\
    $0$ & $1$ & $0$ & $0$ & $1$ & ~$\C\H\C\C\H\C\H\H$~ \\
    $0$ & $1$ & $0$ & $1$ & $1$ & ~$\C\H\C\H\H\C\H\C$~ \\
    $0$ & $1$ & $1$ & $0$ & $1$ & ~$\C\H\H\C\H\C\C\H$~ \\
    $0$ & $1$ & $1$ & $1$ & $0$ & ~$\C\H\H\H\H\C\C\C$~ \\
    $1$ & $0$ & $0$ & $0$ & $0$ & ~$\H\C\C\C\C\H\H\H$~ \\
    $1$ & $0$ & $0$ & $1$ & $1$ & ~$\H\C\C\H\C\H\H\C$~ \\
    $1$ & $0$ & $1$ & $0$ & $1$ & ~$\H\C\H\C\C\H\C\H$~ \\
    $1$ & $0$ & $1$ & $1$ & $1$ & ~$\H\C\H\H\C\H\C\C$~ \\
    $1$ & $1$ & $0$ & $0$ & $0$ & ~$\H\H\C\C\C\C\H\H$~ \\
    $1$ & $1$ & $0$ & $1$ & $1$ & ~$\H\H\C\H\C\C\H\C$~ \\
    $1$ & $1$ & $1$ & $0$ & $0$ & ~$\H\H\H\C\C\C\C\H$~ \\
    $1$ & $1$ & $1$ & $1$ & $0$ & ~$\H\H\H\H\C\C\C\C$~ \\ \hline
    \end{tabular}
    \caption{Correctness of Protocol 2}
    \label{tab:four_variable}
\end{table}

The correctness of Protocol 2 is shown in Table~\ref{tab:four_variable}. 

\begin{enumerate}
    \item The input sequence is given as follows:
    \[
    \underset{x}{\back} \, \underset{y}{\back}\, \underset{z}{\back} \, \underset{w}{\back} \, \underset{\ol{x}}{\back} \, \underset{\ol{y}}{\back}\, \underset{\ol{z}}{\back} \, \underset{\ol{w}}{\back} \,.
    \]
    \item Apply a random cut to the sequence as follows: 
    \[
    \left\langle \; \back\,\back\,\back\,\back\,\back\,\back\,\back\,\back \;\right\rangle. 
    \]
    \item Open all cards. Output $0$ if it is a cyclic shift of $\H\H\H\H\C\C\C\C$, and $1$ if it is a cyclic shift of $\C\C\H\C\H\H\C\H$. 
\end{enumerate}

\subsection{Protocol 3: SCFO Protocol for $x\ol{y} \vee y\ol{z}$}

\begin{table}[t]
    \centering
    \begin{tabular}{|ccc|c|c|}\hline
    ~$x$~ & ~$y$~ & ~$z$~ & ~$x\ol{y} \vee y\ol{z}$~ & ~Input Seq.~\\ \hline\hline
    $0$ & $0$ & $0$ & $0$ & ~$\C\C\C\H\H\H\H\C$~ \\
    $0$ & $0$ & $1$ & $0$ & ~$\C\C\H\H\H\H\C\C$~ \\
    $0$ & $1$ & $0$ & $1$ & ~$\C\H\C\H\H\C\H\C$~ \\
    $0$ & $1$ & $1$ & $0$ & ~$\C\H\H\H\H\C\C\C$~ \\
    $1$ & $0$ & $0$ & $1$ & ~$\H\C\C\H\C\H\H\C$~ \\
    $1$ & $0$ & $1$ & $1$ & ~$\H\C\H\H\C\H\C\C$~ \\
    $1$ & $1$ & $0$ & $1$ & ~$\H\H\C\H\C\C\H\C$~ \\
    $1$ & $1$ & $1$ & $0$ & ~$\H\H\H\H\C\C\C\C$~ \\ \hline
    \end{tabular}
    \caption{Correctness of Protocol 3}
    \label{tab:if_then_else}
\end{table}

By executing Protocol 2 with $w = 1$, 
we obtain a new SCFO protocol for $x\ol{y} \vee y\ol{z}$. This function outputs $x$ if $y = 0$ and $\ol{z}$ if $y = 1$. 

\begin{enumerate}
    \item The input sequence is given as follows:
    \[
    \underset{x}{\back} \, \underset{y}{\back}\, \underset{z}{\back} \, \underset{1}{\back} \, \underset{\ol{x}}{\back} \, \underset{\ol{y}}{\back}\, \underset{\ol{z}}{\back} \, \underset{0}{\back} \,.
    \]
    \item Apply a random cut to the sequence as follows: 
    \[
    \left\langle \; \back\,\back\,\back\,\back\,\back\,\back\,\back\,\back \;\right\rangle. 
    \]
    \item Open all cards. Output $0$ if it is a cyclic shift of $\H\H\H\H\C\C\C\C$, and $1$ if it is a cyclic shift of $\C\C\H\C\H\H\C\H$. 
\end{enumerate}

The correctness of Protocol 3 is shown in Table~\ref{tab:if_then_else}. 

\section{Conclusion}

In this paper, we proposed three new SCFO protocols in addition to the three existing ones.  
Determining for which Boolean functions $f: \bin^n \ra \bin$ an SCFO protocol can be constructed remains a significant open problem in the field of card-based cryptography.  
It should be noted that our proposed protocol for $x \oplus y \oplus z$ employs two pairs of $(x, \ol{x})$, which distinguishes it from the other SCFO protocols presented.  
It is also an intriguing research direction to explore the feasibility of constructing SCFO protocols both in the setting where multiple input pairs are allowed and in the setting where they are not.

\bibliographystyle{abbrv}
\bibliography{card}

\end{document}